\documentclass[11pt]{article}
\usepackage{amssymb,latexsym,amsmath}
\usepackage[dvips]{graphicx}
\usepackage{cite}
\newtheorem{theo}{Theorem}

\newtheorem{lemm}[theo]{Lemma}

\newtheorem{prop}[theo]{Proposition}
\def\nn{\nonumber}

\def\qdots{\mathinner{\mkern1mu\raise1pt\vbox{\kern7pt\hbox{.}}\mkern2mu
 \raise4pt\hbox{.}\mkern2mu\raise7pt\hbox{.}\mkern1mu}}

\newcommand{\q}{{\hat q}}
\newcommand{\p}{{\hat p}}
\newcommand{\hH}{{\hat H}}

\renewcommand{\atop}[2]{\genfrac{}{}{0pt}{}{#1}{#2}}

\def\mybox{\hfill$\Box$}
\begin{document}
\begin{center}
{\Large \bf
Analytically solvable Hamiltonians for quantum systems\\[2mm] 
with a nearest neighbour interaction 
}\\[5mm]
{\bf G.~Regniers\footnote{E-mail: Gilles.Regniers@UGent.be}, }
{\bf and J.\ Van der Jeugt}\footnote{E-mail:
Joris.VanderJeugt@UGent.be}\\[1mm]
Department of Applied Mathematics and Computer Science,
Ghent University,\\
Krijgslaan 281-S9, B-9000 Gent, Belgium.
\end{center}

\vskip 10mm
\noindent
Short title: solvable quantum Hamiltonians



\begin{abstract}
We consider quantum systems consisting of a linear chain of $n$ harmonic oscillators
coupled by a nearest neighbour interaction of the form $-\q_r \q_{r+1}$ ($\q_r$ refers to
the position of the $r$th oscillator). 
In principle, such systems are always numerically solvable and involve the eigenvalues of the
interaction matrix.
In this paper, we investigate when such a system is analytically solvable, i.e.\
when the eigenvalues and eigenvectors of the interaction matrix have analytically
closed expressions.
This is the case when the interaction matrix coincides with the Jacobi matrix of
a system of discrete orthogonal polynomials.
Our study of possible systems leads to three new analytically solvable
Hamiltonians: with a Krawtchouk interaction, a Hahn interaction or a $q$-Krawtchouk interaction.
For each of these cases, we give the spectrum of the Hamiltonian (in analytic form) and 
discuss some typical properties of the spectra.
\end{abstract}

%

\setcounter{equation}{0}
\section{Introduction} \label{sec1}

In classical mechanics, one-dimensional systems (or lattices) consisting of mass points
with some nearest neighbour interaction have a long history. A typical system is a lattice
of $n$ particles with masses $m_1, m_2,\ldots, m_n$, and a harmonic coupling with spring
constants $\kappa_1, \kappa_2,\ldots, \kappa_{n-1}$ leading to the Hamiltonian
\begin{equation}
H(p,x)= \sum_{r=1}^n \frac{p_r^2}{2m_r} + \sum_{r=1}^{n-1} \frac{\kappa_r}{2}(x_j-x_{j+1})^2.
\end{equation}
Such classical systems (or variations, with an infinite number of mass points, or with
various boundary conditions) were already considered by Schr\"odinger~\cite{Schrodinger}.
The equations of motion of such a system can be solved by (numerically) diagonalizing
the interaction matrix (the eigenvalues of which yield the normal modes of the system).
Alternatively, the system can be solved using orthogonal polynomials whose
recurrence relations are derived from the equations of motion~\cite{Christian,Law,Mokole}.
In that case, the normal modes are obtained from the zeros of the $n$th orthogonal polynomial.

One-dimensional systems with a different type of nearest neighbour interaction received a lot of
attention, especially those that are still exactly solvable. Among the most famous, we mention
the Toda system~\cite{Toda} and the Calogero-Sutherland-Moser models~\cite{Calogero,Sutherland,Moser}.
In this context, the emphasis was shifted from physics to mathematical aspects such
as integrability and the underlying algebraic structures. 

Also the quantum versions of many of these systems or models were investigated from various points
of view during the last decades. 
Quantum Calogero-Moser systems for any root system were studied by Olshanetsky and 
Perelomov~\cite{Olshanetsky}; for a review, see~\cite{Corrigan}.
In such quantum systems, the emphasis -- from the physics point of view -- is on a construction
of ground wave states, formulae for the excitation spectrum, a description of stationary states, etc.
Several other quantum systems with a nearest neighbour interaction closely related to Calogero-Sutherland-Moser
models were explored, see~\cite{Karimipour,Jain,Basu,Auberson} to cite a few.

In the present paper we consider yet another quantum system given in equation~\eqref{H-gen},
consisting of a one-dimensional
chain of particles with a certain nearest neighbour interaction, which is quadratic in 
the position operators. 
Our emphasis is on the investigation of analytical solvability of the quantum system (i.e.\
on obtaining analytically closed expressions of the spectrum of the Hamiltonian and of
its eigenstates, the stationary states of the system).
In such a context, the physical significance of the interaction introduced here is less clear:
it can, in a sense, be considered as a deviation from a vibrating quantum system.

To introduce our system, let us first consider one of the most common
quantum systems, consisting of
a chain of harmonic oscillators coupled by some nearest neighbour 
interaction~\cite{Plenio,Cohen,LSV06,LSV08}. 
In this popular model the particles are described as identical harmonic oscillators 
which are moreover coupled by springs obeying Hooke's law. 
Then the Hamiltonian of the system is given by:
\begin{equation}
\hH_h=\sum_{r=1}^{n} \Big( \frac{\hat{p}_r^2}{2m}
+ \frac{m\omega^2}{2} \hat{q}_r^2 \Big) + \sum_{r=0}^n \frac{cm}{2}(\hat{q}_r-\hat{q}_{r+1})^2 ,
\label{H1}
\end{equation}
where $\hat{q}_0=\hat{q}_{n+1}\equiv 0$ (fixed wall boundary conditions).
In other words, the quantum system consists of a string or chain of $n$ identical 
harmonic oscillators, each having the same mass $m$ and natural frequency $\omega$.
The position and momentum operators for the $r$th oscillator are given by $\q_r$ and $\p_r$;
more precisely $\q_r$ measures the displacement of the $r$th mass point with respect to its
equilibrium position.
The last term in~\eqref{H1} represents the nearest neighbour coupling by
means of ``springs'', with a coupling strength~$c$ ($c\geq0$). For $c=0$ one is simply dealing
with a set of identical uncoupled harmonic oscillators.

We shall consider~\eqref{H1}, and the other quantum system being studied here,
in the case of canonical quantization, i.e.\ the $\q_r$ and $\p_r$
are self-adjoint operators ($\q_r^\dagger=\q_r$ and $\p_r^\dagger=p_r$) satisfying the commutation
relations
\begin{equation}
[\q_r, \q_s ] =0,\qquad [\p_r, \p_s ] =0, \qquad [\q_r, \p_s ] =i\hbar\delta_{rs}
\qquad(r,s=1,\ldots,n).
\end{equation}
It is well known that the Hamiltonian~\eqref{H1} is completely solvable (see also next section). In fact,
it is analytically solvable in the sense that one has an analytically closed expression
for the eigenvalues of~\eqref{H1}.

Note that~\eqref{H1} can be rewritten in the following form (with $\Tilde\omega^2=\omega^2+2c$)
\begin{equation}
\hH_h=\sum_{r=1}^{n} \Big( \frac{\hat{p}_r^2}{2m}
+ \frac{m\Tilde\omega^2}{2} \hat{q}_r^2 \Big) - cm\sum_{r=1}^{n-1} \hat{q}_r \hat{q}_{r+1}.
\label{H1-tilde}
\end{equation}
The fact that the nearest neighbour interaction is the same everywhere in the chain,
thus independent of the position~$r$, implies that all coefficients of $\hat{q}_r \hat{q}_{r+1}$ 
are the same.

In the present paper, we shall study deviations of~\eqref{H1-tilde}, where the nearest neighbour
interaction is not constant in the chain, but depends on~$r$.
The Hamiltonian of such a more general system is assumed to be of the following form:
\begin{equation}
\hH =\sum_{r=1}^{n} \Big( \frac{\hat{p}_r^2}{2m}
+ \frac{m\omega^2}{2} \hat{q}_r^2 \Big) - \frac{cm}{2}\sum_{r=1}^{n-1} \gamma_r \hat{q}_r \hat{q}_{r+1}.
\label{H-gen}
\end{equation}
The physical interpretation of such a system is not obvious.
First of all, in order to have a physical meaning, the interaction matrix related to~\eqref{H-gen}
should still be positive definite (see next section).
Second, the interaction is in general no longer harmonic. 
It could be seen as a chain still consisting of identical oscillators but with a
quadratic nearest neighbour interaction of the form $-\hat{q}_r \hat{q}_{r+1}$.
This interaction is not homogeneous in the chain but 
depends on the location $r$ in the linear system. 
For a more general context in which quantum systems of the form~\eqref{H-gen} appear as
a special case, see the notion of ``harmonic systems on general lattices'' in~\cite{Cramer,Amico}
in the study of entanglement in many-body systems. 

The purpose of this paper is to study the spectrum of Hamiltonians of the form~\eqref{H-gen}. 
Such Hamiltonians are always numerically solvable, i.e.\ the complete spectrum can be
described using the eigenvalues of the interaction matrix (see next section).
The contribution of this paper is to investigate those cases for which~$\hH$ is analytically solvable
(i.e.\ when the eigenvalues of the interaction matrix are known in analytically closed form).
Some examples of Hamiltonians for which this is the case are:
\begin{align*}
\hH_K&=\sum_{r=1}^{n} \Big( \frac{\hat{p}_r^2}{2m}
+ \frac{m\omega^2}{2} \hat{q}_r^2 \Big) - \frac{cm}{2} \sum_{r=1}^{n-1} \sqrt{r(n-r)}\,
\hat{q}_r\hat{q}_{r+1} ,\\
\hat{H}_Q^{(1)}&=\sum_{r=1}^{n} \Big( \frac{\hat{p}_r^2}{2m}
+ \frac{m\omega^2}{2} \hat{q}_r^2 \Big) - \frac{cm}{2} \sum_{r=1}^{n-1} 
\frac{\sqrt{(n-r)(n+r+1)}}{2}\,
\hat{q}_r\hat{q}_{r+1} ,\\
\hH_{Kq}&=\sum_{r=1}^{n} \Big( \frac{\hat{p}_r^2}{2m}
+ \frac{m\omega^2}{2} \hat{q}_r^2 \Big) - \frac{cm}{2} \sum_{r=1}^{n-1} 
2\sqrt{q^{r+1-2n}(1-q^r)(1-q^{n-r})}\,
\hat{q}_r\hat{q}_{r+1} ,
\end{align*}
where $q$ is some positive parameter.

In the following section, we shall consider~\eqref{H1} again, and describe a method to solve
this Hamiltonian. This method can best be formulated in terms of an 
interaction matrix $M$~\cite{Iachello,Karimipour},
and we shall determine the conditions for $M$ to be analytically solvable. 
In particular, systems of the form~\eqref{H-gen} correspond to tridiagonal interaction matrices with
a constant diagonal. 
As we shall see, this leads us to the area of discrete orthogonal polynomials.
We have investigated the families of discrete orthogonal polynomials that lead to solutions.
These are described in the following sections: an interaction based upon Krawtchouk polynomials,
on Hahn polynomials, or on dual $q$-Krawtchouk polynomials.
For each of these cases, we give the corresponding Hamiltonian and its solution.
In section~\ref{sec6} we also briefly describe some interesting features of the energy levels
of these Hamiltonians.

\section{General method} \label{sec2}

The Hamiltonian~\eqref{H1} can be written in matrix form as follows:
\begin{equation}
\hH_h=\frac{1}{2m} \left(\begin{array}{ccc} \p_1^\dagger & \cdots & \p_n^\dagger \end{array}\right)
\left(\begin{array}{c} \p_1 \\ \vdots \\ \p_n \end{array}\right) + 
\frac{m}{2} \left(\begin{array}{ccc} \q_1^\dagger & \cdots & \q_n^\dagger \end{array}\right) A_h 
\left(\begin{array}{c} \q_1 \\ \vdots \\ \q_n \end{array}\right),
\end{equation}
where $A_h$ is a symmetric tridiagonal matrix of the form
\begin{equation}
A_h = \omega^2 I + c M_h,
\end{equation}
with $I$ the $n\times n$ identity matrix and 
\begin{equation}
M_h=
\left(\begin{array}{cccccc}
2 & -1 & 0 & \cdots & 0 & 0 \\
-1 & 2 & -1 & \cdots & 0 & 0 \\
0 & -1 & 2 & \cdots & 0 & 0 \\
\vdots & \vdots & \vdots & \ddots & \vdots & \vdots \\
0 & 0 & 0 & \cdots & 2 & -1 \\
0 & 0 & 0 & \cdots & -1 & 2
\end{array} \right).
\label{Mh}
\end{equation}
The fact that the interaction is only between nearest neighbours is reflected by the tridiagonal
form of $M_h$ (i.e.\ nonzero entries only on the diagonal, the subdiagonal and the superdiagonal).
We shall refer to $M_h$ as the ``interaction matrix''~\cite{Iachello,Karimipour}.

Let us for a while consider quantum systems described by a Hamiltonian with a more general 
interaction matrix $M$:
\begin{equation}
\hat{H}=\frac{1}{2m} \left(\begin{array}{ccc} \p_1^\dagger & \cdots & \p_n^\dagger \end{array}\right)
\left(\begin{array}{c} \p_1 \\ \vdots \\ \p_n \end{array}\right) + 
\frac{m}{2} \left(\begin{array}{ccc} \q_1^\dagger & \cdots & \q_n^\dagger \end{array}\right) (\omega^2 I + c M) 
\left(\begin{array}{c} \q_1 \\ \vdots \\ \q_n \end{array}\right).
\label{H}
\end{equation}
In~\eqref{H}, $M$ is a {\em real and symmetric} matrix. 
In order to be physically meaningful, $\omega^2 I + c M$ should be a positive definite 
matrix~\cite{Cramer,Amico}.
A general method to deal with such Hamiltonians was described in~\cite[section~2.1]{Cramer}.
Since $M$ is real and symmetric, the spectral theorem~\cite{Golub} implies
\begin{equation}
M = UDU^T
\label{decomp}
\end{equation}
where
\begin{align}
& D = \hbox{diag}(\lambda_1,\lambda_2,\ldots,\lambda_n), \label{D}\\
& UU^T = U^TU=I. \label{UUT}
\end{align}
The entries of the diagonal matrix $D$ are the (real) eigenvalues $\lambda_i$ of $M$, in some order, and
the columns of the real orthogonal matrix $U$ are eigenvectors of $M$ (in the same order); $U^T$ stands for
the transpose of $U$.

Introducing new operators (the so-called normal coordinates and momenta) as follows:
\begin{equation}
\left(\begin{array}{c} \hat Q_1\\ \vdots \\ \hat Q_n\end{array}\right) = U^T
\left(\begin{array}{c} \q_1 \\ \vdots \\ \q_n \end{array}\right),
\qquad
\left(\begin{array}{c} \hat P_1\\ \vdots \\ \hat P_n\end{array}\right) = U^T
\left(\begin{array}{c} \p_1 \\ \vdots \\ \p_n \end{array}\right),
\label{tf}
\end{equation}
the Hamiltonian~\eqref{H} reads
\begin{align}
\hat{H}& =\frac{1}{2m} \left(\begin{array}{ccc} \hat P_1^\dagger & \cdots & \hat P_n^\dagger \end{array}\right)
\left(\begin{array}{c} \hat P_1 \\ \vdots \\ \hat P_n \end{array}\right) + 
\frac{m}{2} \left(\begin{array}{ccc} \hat Q_1^\dagger & \cdots & \hat Q_n^\dagger \end{array}\right) (\omega^2 I + c D) 
\left(\begin{array}{c} \hat Q_1 \\ \vdots \\ \hat Q_n \end{array}\right) \nn \\
& = \frac{1}{2m} \sum_{j=1}^n {\hat P}_j^2 + \frac{m}{2} \sum_{j=1}^n (\omega^2+c\lambda_j) {\hat Q}_j^2.
\label{HPQ}
\end{align}
By the transformation~\eqref{tf}, the new operators also satisfy the canonical commutation relations:
\begin{equation}
[\hat Q_j, \hat Q_k ] =0,\qquad [\hat P_j, \hat P_k ] =0, \qquad [\hat Q_j, \hat P_k ] =i\hbar\delta_{jk}
\qquad(j,k=1,\ldots,n).
\end{equation}
In~\eqref{HPQ}, the values of $\omega^2+c\lambda_j$ are all positive since
the interaction matrix $\omega^2 I + c M$ is assumed to be positive definite.
So one can introduce
\begin{equation}
\omega_j = \sqrt{\omega^2+c \lambda_j}.
\end{equation}
Then we can write
\begin{equation}
\hat H = \frac{1}{2m} \sum_{j=1}^n {\hat P}_j^2 + \frac{m}{2} \sum_{j=1}^n \omega_j^2 {\hat Q}_j^2.
\end{equation}
This expression is just like the Hamiltonian of an $n$-dimensional non-isotropic oscillator, so we can use the
commonly known method for its solution~\cite{Moshinsky,Wybourne}. Introducing boson operators
\begin{equation}
a^\pm_j = \sqrt{\frac{m\omega_j}{2\hbar}} \hat Q_j \mp \frac{i}{\sqrt{2\hbar m \omega_j}} \hat P_j,
\end{equation}
these satisfy
\begin{equation}
[a_j^-,a_k^-]=[a_j^+,a_k^+]=0,\qquad [a_j^-,a_k^+]=\delta_{jk}, \qquad(j,k=1,\ldots,n)
\end{equation}
and $\hat H$ can be written as
\begin{equation}
\hat H = \sum_{j=1}^n \frac{\hbar\omega_j}{2} \{ a_j^+, a_j^-\} = 
\sum_{j=1}^n \frac{\hbar\omega_j}{2} (2 a_j^+ a_j^- +1).
\end{equation}
Furthermore,
\begin{equation}
[ \hat H, a_j^\pm ] = \pm \hbar\omega_j \, a_j^\pm \qquad(j=1,\ldots,n).
\end{equation}
So if we assume that there is a lowest $\hat H$-eigenvalue (lowest energy), say for the state $|0\rangle$, then we
have the usual $n$-boson Fock space in which the action of $\hat H$ is diagonal. The vacuum vector $|0\rangle$
satisfies
\begin{equation}
\langle 0|0 \rangle = 1, \qquad a_j^- |0\rangle = 0;
\end{equation}
the other (orthogonal and normalized) basis vectors are then defined by
\begin{equation}
| k_1, \ldots, k_n \rangle = 
\frac{ (a_1^+)^{k_1} \ldots (a_n^+)^{k_n} }{\sqrt{k_1! \ldots k_n!} } |0\rangle,\qquad
(k_j=0,1,\ldots).
\end{equation}
The spectrum of $\hat H$ is now determined by
\begin{equation}
\hat H  | k_1, \ldots, k_n \rangle = \sum_{j=1}^n \hbar\omega_j (k_j+\frac{1}{2}) \;
| k_1, \ldots, k_n \rangle.
\label{specH}
\end{equation}

This analysis is well known, and it seems to indicate that a Hamiltonian of the form~\eqref{H}
with a general interaction matrix $M$ is exactly solvable as a quantum system. 
Note, however, that the solution we have described involves also a {\em numerical process}, namely the 
determination of the eigenvalues and eigenvectors of $M$ in~\eqref{D} and~\eqref{UUT}.
We shall say that the Hamiltonian $\hat H$ is {\em analytically solvable} if we have an 
analytically closed expression for the eigenvalues and eigenvectors of $M$, for arbitrary $n$.

One example of an analytically solvable Hamiltonian is~\eqref{H1}, with interaction matrix $M_h$
given by~\eqref{Mh}.
In this case, the decomposition~\eqref{decomp} is determined by~\cite{Plenio,LSV08}
\begin{equation}
U=\frac{\sqrt{2}}{\sqrt{n+1}} \left( \sin( \frac{ij\pi}{n+1} ) \right)_{1\leq i,j\leq n}
\end{equation}
and the eigenvalues in~\eqref{D} by
\begin{equation}
\lambda_j = 2 - 2\cos( \frac{j\pi}{n+1} ).
\end{equation}
So we have
\begin{equation}
\omega_j^2 = \omega^2+2c-2c\cos( \frac{j\pi}{n+1} ) = \omega^2 + 4c \sin^2( \frac{j\pi}{2(n+1)} ).
\label{omega-a}
\end{equation}

Apparently, there are not so many examples of analytically solvable Hamiltonians of the type~\eqref{H}
known in the literature. One paper dealing with this problem (and closely related ones)
is~\cite{Iachello}. In that paper,
some examples of analytically solvable Hamiltonians are given. 

In the current paper, we present some new examples. 
Since we are studying Hamiltonians of the form~\eqref{H-gen}, we are dealing with tridiagonal
interaction matrices $M$ with constant entries on the diagonal.
For tridiagonal matrices, an explicit spectral decomposition~\eqref{decomp}
can be found by relating these matrices to Jacobi matrices of discrete orthogonal polynomials.
So it is natural to look for new examples in that area.
We shall first describe the example of ``Krawtchouk interaction'', and then indicate 
how to find other examples.

\section{Krawtchouk interaction} \label{sec3}

In this section, let us first collect some (known) properties of Krawtchouk polynomials,
and then use these to describe the spectrum of a Hamiltonian with a Krawtchouk interaction term.
For a list of hypergeometric orthogonal polynomials, see~\cite{Koekoek} or~\cite{Ismail}.

\subsection{Krawtchouk polynomials}
For a fixed positive integer parameter $N$ and a real parameter $p$ ($0<p<1$), 
the Krawtchouk polynomial of degree $i$ ($i=0,1,\ldots,N$) in the variable $x$ is defined by~\cite{Koekoek,Ismail,Suslov}
\begin{equation}
K_i(x)\equiv K_i(x; p,N) = \mbox{$_2F_1$} \left( \atop{-x,-i}{-N} ; \frac{1}{p} \right).
\label{defK}
\end{equation}
Herein, ${}_2F_1$ is the usual Gauss hypergeometric series~\cite{Slater}
\begin{equation}
\mbox{$_2F_1$} \left( \atop{a,b}{c} ; z\right) = \sum_{k=0}^\infty \frac{(a)_k (b)_k}{(c)_k} \frac{z^k}{k!}.
\label{def2F1}
\end{equation}
In~\eqref{defK}, the series is terminating because one of the numerator parameters is a negative integer.
Note that in~\eqref{def2F1} we use the notation of the raising factorial, which can also be rewritten by
means of a (generalized) binomial coefficient:
\[
(a)_k = a (a+1) \cdots (a+k-1) = (-1)^k \binom{-a}{k} k!
\]
The Krawtchouk polynomials satisfy a discrete orthogonality relation of the form
\begin{equation}
\sum_{x=0}^N w(x) K_i(x) K_j(x) = h_i \delta_{ij},
\label{orth-K}
\end{equation} 
where $w(x)$ is a weight function in $x$ and $h_i$ is a function depending on $i$:
\begin{equation}
w(x) = \binom{N}{x} \, p^x \, (1-p)^{N-x}\qquad (x=0,1,\ldots,N); \qquad\qquad
h_i = \frac{1}{\binom{N}{i}} \left( \frac{1-p}{p} \right)^i.
\end{equation}
Recall that the recurrence relation for Krawtchouk polynomials is given by
\begin{equation}
-x K_i(x)  =  i(1-p) \, K_{i-1}(x)- \bigl[ p(N-i)+i(1-p) \bigr] \, K_i(x)
+ \, p(N-i) \, K_{i+1}(x). \label{kraw_rec1}
\end{equation}
For future purposes we will however be interested in an orthonormality condition, so we define the orthonormal Krawtchouk polynomials by
\begin{equation}
\label{Krawtchouk}
\Tilde K_i(x)\equiv \Tilde K_i(x; p,N) = \frac{\sqrt{w(x)} \, K_i(x)}{\sqrt{h_i}},  \qquad i = 0,1,2, \ldots, N.
\end{equation}
Now we can state the following property:
\begin{lemm} \label{kraw_EF}
Let $M_K$ be the tridiagonal $(N+1)\times(N+1)$-matrix
\begin{equation}
\label{MK}
M_K= \left( \begin{array}{ccccc}
             F_0 & -E_1  &    0   &        &      \\
            -E_1 &  F_1  &  -E_2  & \ddots &      \\
              0  & -E_2  &   F_2  & \ddots &  0   \\
                 &\ddots & \ddots & \ddots & -E_N \\
                 &       &    0   &  -E_N  &  F_N
          \end{array} \right),
\end{equation}
where
\begin{equation} 
\label{kraw_E}
E_i = \sqrt{p(1-p)} \sqrt{i(N-i+1)}, \qquad F_i = Np + (1-2p)i,
\end{equation}
and let $U$ be the $(N+1)\times(N+1)$-matrix with matrix elements
\begin{equation}
U_{ij} = \Tilde K_i(j) = \left[ \binom{N}{i}\binom{N}{j}p^{i+j}(1-p)^{N-i-j}\right]^{1/2} \;
\sum_{k=0}^{\min(i,j)} \frac{\binom{i}{k}\binom{j}{k}}{\binom{N}{k}} (-\frac{1}{p})^k,
\end{equation}
where $i,j=0,1,\ldots,N$. 
Then
\begin{equation}
UU^T = U^TU=I \qquad\hbox{and} \qquad M_K = UDU^T
\end{equation}
where $D = \hbox{diag}(0,1,2\ldots,N)$.
\end{lemm}

\noindent {\bf Proof.}
We have that
\[
(U U^T)_{ij} = \sum_{k=0}^N U_{ik} U_{jk} = \sum_{k=0}^N \Tilde K_i(k) \Tilde K_j(k) = \delta_{ij}
\]
by the orthogonality relations~\eqref{orth-K}.
So $U U^T=I$, hence $U^T$ is the inverse of $U$ and thus also $U^TU=I$.
Furthermore, notice that
\[
E_i = i(1-p) \sqrt{\frac{h_{i-1}}{h_i}}
\qquad\hbox{and}\qquad
E_{i+1} = p(N-i) \sqrt{\frac{h_{i+1}}{h_i}}.
\]
Equation~\eqref{kraw_rec1} can then be rewritten as a recurrence relation for the orthonormal 
Krawtchouk polynomials $\Tilde K_i(x)$:
\begin{equation}
\label{kraw_rec2}
x \Tilde K_i(x) = - E_i \,\Tilde K_{i-1}(x)
           + F_i \,\Tilde K_i(x) 
           - E_{i+1} \,\Tilde K_{i+1}(x).
\end{equation}
Then we have, using $E_0=E_{N+1}=0$,
\begin{align}
(M_K U)_{ij} & = \sum_{k=0}^N (M_K)_{ik} U_{kj} \nn\\
& = -E_i \;\Tilde K_{i-1}(j) +F_i \;\Tilde K_{i}(j) -E_{i+1} \;\Tilde K_{i+1}(j) \nn\\
& = j\;\Tilde K_i(j) = (U D)_{ij}, \nn
\end{align}
so $M_K U= UD$ or $M_K = UDU^T$.
\mybox

So we now have a good candidate interaction matrix $M_K$. 
In order to describe systems of the form~\eqref{H-gen}, however, the diagonal entries $F_{i}$ of $M_K$
should be constants (i.e.\ independent of $i$).
We see from~\eqref{kraw_E} that this is the case for $p=1/2$.
So this leads to a new analytically solvable Hamiltonian of the form~\eqref{H-gen}.

\subsection{Hamiltonian with Krawtchouk interaction}

Consider a linear chain of $n$ identical harmonic oscillators, with a nearest neighbour 
interaction that is given by
\begin{equation}
\hH_K=\sum_{r=1}^{n} \Big( \frac{\hat{p}_r^2}{2m}
+ \frac{m\omega^2}{2} \hat{q}_r^2 \Big) - \frac{cm}{2} \sum_{r=1}^{n-1} \sqrt{r(n-r)}\;
\hat{q}_r\hat{q}_{r+1} .
\label{HK}
\end{equation}
We shall refer to the interaction term as ``Krawtchouk interaction''.
The purpose is to find the analytic solution for the spectrum of $H_K$.
It is easy to see that this Hamiltonian can be written in matrix form, like~\eqref{H}:
\begin{equation}
\hH_K=\frac{1}{2m} \left(\begin{array}{ccc} \p_1^\dagger & \cdots & \p_n^\dagger \end{array}\right)
\left(\begin{array}{c} \p_1 \\ \vdots \\ \p_n \end{array}\right) + 
\frac{m}{2} \left(\begin{array}{ccc} \q_1^\dagger & \cdots & \q_n^\dagger \end{array}\right) 
\big((\omega^2-\frac{c(n-1)}{2}) I + c M_K\big) 
\left(\begin{array}{c} \q_1 \\ \vdots \\ \q_n \end{array}\right),
\label{HKm}
\end{equation}
where $M_K$ is the matrix~\eqref{MK} with $N=n-1$ and $p=1/2$.
But for the matrix $M_K$ we have an explicit spectral decomposition, given in Lemma~\ref{kraw_EF}, 
and the eigenvalues of $M_K$ are given by $0,1,\ldots,n-1$. 
Using this result, and following the general procedure described in section~\ref{sec2}, one
introduces here the following quantities:
\begin{equation}
\omega_j = \sqrt{ \omega^2-\frac{c(n-1)}{2} + c(j-1)  } = 
\sqrt{ \omega^2-\frac{c}{2}(n-2j+1)  }, \qquad (j=1,2,\ldots,n).
\label{omegaK}
\end{equation}
The interaction matrix $(\omega^2-\frac{c(n-1)}{2}) I + c M_K$ is positive definite
if all quantities under the square root symbol are positive. 
Since $c$ (and $\omega^2$) is positive, $\omega^2-\frac{c(n-1)}{2} + c(j-1)$ ($j=1,2,\ldots,n$)
is an increasing sequence as $j$ increases. So this condition leads to $c<2\omega^2/(n-1)$,
or the ``coupling strength'' should be sufficiently small.
Now we have:
\begin{prop}
The Hamiltonian $\hat H_K$ given by~\eqref{HK} is analytically solvable.
The explicit spectrum of $H_K$ follows from~\eqref{specH}:
\begin{equation}
\hat H_K  | k_1, \ldots, k_n \rangle = \sum_{j=1}^n \hbar\omega_j (k_j+\frac{1}{2}) \;
| k_1, \ldots, k_n \rangle,
\label{specHK}
\end{equation}
where the constants $\omega_j$ are given by $\omega_j =\sqrt{ \omega^2-c(n-2j+1)/2}$.
\end{prop}

Finally, notice that the interaction term in~\eqref{HK} is invariant under the reflection $r\rightarrow n-r$.

\subsection{Remark}
It is clear that the general procedure worked out here for the Krawtchouk polynomials works
in general for discrete orthogonal polynomials. 
So in order to find other interesting examples, one can go through the list of 
discrete orthogonal polynomials~\cite{Koekoek,Ismail,Suslov} and their $q$-analogues.
The basic restriction, in order to have Hamiltonians of the form~\eqref{H-gen},
is that the diagonal elements in the interaction matrix are constant (for specific
values of the parameters).
An investigation of this restriction has shown that, apart from the Krawtchouk
polynomials with $p=1/2$, there are only the following cases to be considered:
the Hahn polynomials with $\beta=\alpha$
and the dual $q$-Krawtchouk polynomials with $\bar c=-1$.
We shall now study these cases and the corresponding Hamiltonians.

\section{Hahn interaction} \label{sec4}

\subsection{Hahn polynomials}

Hahn polynomials are again a class of discrete orthogonal polynomials $Q_i(x;\alpha,\beta,N)$, characterized by a 
positive integer parameter $N$ and two real parameters $\alpha$ and $\beta$ 
(for orthogonality, one should have $\alpha>-1$ and $\beta>-1$, or $\alpha<-N$ and $\beta<-N$).
The Hahn polynomial of degree $i$ ($i=0,1,\ldots,N$) in the variable $x$ is defined by~\cite{Koekoek,Ismail}
\begin{equation}
Q_i(x) \equiv Q_i(x;\alpha,\beta,N) = {\;}_3F_2 \left( \atop{-i,i+\alpha+\beta+1,-x}{\alpha+1,-N} ; 1 \right),
\label{defQ}
\end{equation}
where ${\;}_3F_2$ is the generalized hypergeometric series (which is terminating here due to the
numerator parameter $-i$). 
The Hahn polynomials satisfy a discrete orthogonality relation:
\begin{equation}
\sum_{x=0}^N w(x) Q_i(x) Q_j(x) = h_i \delta_{ij},
\label{orth-Q}
\end{equation} 
where
\begin{align*}
w(x) &= \binom{\alpha+x}{x} \binom{N+\beta-x}{N-x} \quad (x=0,1,\ldots,N); \\
h_i &= \frac{i!(N-i)!}{N!^2}\frac{(i+\alpha+\beta+1)_{N+1}(\beta+1)_i}{(2i+\alpha+\beta+1)(\alpha+1)_i}.
\end{align*}
The recurrence relation for Hahn polynomials is given by
\begin{equation}
-x Q_i(x)  =  A_i \, Q_{i+1}(x)- (A_i+C_i) \, Q_i(x)
+ C_i \, Q_{i-1}(x), \label{Q-rec1}
\end{equation}
where
\[
A_i= \frac{(i+\alpha+\beta+1)(i+\alpha+1)(N-i)}{(2i+\alpha+\beta+1)(2i+\alpha+\beta+2)}, \quad
C_i=\frac{i(i+\alpha+\beta+N+1)(i+\beta)}{(2i+\alpha+\beta)(2i+\alpha+\beta+1)}.
\]
The orthonormal Hahn polynomials are defined by
\begin{equation}
\label{Q-orth}
\Tilde Q_i(x) = \frac{\sqrt{w(x)} \, Q_i(x)}{\sqrt{h_i}},  \qquad i = 0,1,2, \ldots, N.
\end{equation}
Then we have a similar lemma as Lemma~\ref{kraw_EF}:
\begin{lemm} \label{Q-lemma}
Let $M_Q$ be the tridiagonal $(N+1)\times(N+1)$-matrix
\begin{equation}
\label{MQ}
M_Q= \left( \begin{array}{ccccc}
             F_0 & -E_1  &    0   &        &      \\
            -E_1 &  F_1  &  -E_2  & \ddots &      \\
              0  & -E_2  &   F_2  & \ddots &  0   \\
                 &\ddots & \ddots & \ddots & -E_N \\
                 &       &    0   &  -E_N  &  F_N
          \end{array} \right),
\end{equation}
where
\begin{align} 
E_i &=\sqrt{ \frac{i \, (i+\alpha) \, (i+\beta) \, (i+\alpha+\beta) \, (i+\alpha+\beta+N+1) \, (N-i+1)}
                  {(2i+\alpha+\beta)^2(2i+\alpha+\beta-1)(2i+\alpha+\beta+1)} },\nn\\
F_i &= \frac{N}{2} + \frac{(\alpha-\beta) \bigl[ (\alpha+\beta) \, (N-2i) - 2i(i+1) \bigr]}
                         {2(2i+\alpha+\beta) \, (2i+\alpha+\beta+2)},
\label{Q_EF}
\end{align}
and let $U$ be the $(N+1)\times(N+1)$-matrix with matrix elements
\begin{equation}
U_{ij} = \Tilde Q_i(j) 
\end{equation}
where $i,j=0,1,\ldots,N$. 
Then
\begin{equation}
UU^T = U^TU=I \qquad\hbox{and} \qquad M_Q = UDU^T
\end{equation}
where $D = \hbox{diag}(0,1,2\ldots,N)$.
\end{lemm}

\noindent {\bf Proof.}
The proof is essentially the same as that of Lemma~\ref{kraw_EF}:
\[
(U U^T)_{ij} = \sum_{k=0}^N U_{ik} U_{jk} = \sum_{k=0}^N \Tilde Q_i(k) \Tilde Q_j(k) = \delta_{ij}
\]
by the orthogonality relations~\eqref{orth-Q}, hence $UU^T = U^TU=I$.
Furthermore, equation~\eqref{Q-rec1} can then be rewritten as a recurrence relation for the orthonormal 
Hahn polynomials $\Tilde Q_i(x)$:
\begin{equation}
\label{Q_rec2}
x \Tilde Q_i(x) = - E_i \,\Tilde Q_{i-1}(x)
           + F_i \,\Tilde Q_i(x) 
           - E_{i+1} \,\Tilde Q_{i+1}(x),
\end{equation}
and this implies $M_Q U= UD$ or $M_Q = UDU^T$.
\mybox

It remains to be determined when the diagonal of~\eqref{MQ} is constant, in other words when $F_i$ is 
independent of~$i$. 
Following~\eqref{Q_EF}, this happens when $\beta=\alpha$.
Note that in that case the parameter should satisfy $\alpha>-1$ or $\alpha<-N$.
It is worthwhile mentioning that there are two other cases where the
diagonal of~\eqref{MQ} is ``almost constant'':
\begin{itemize}
\item
When $\beta=-\alpha$ (with $-1<\alpha<1$) one finds that all $F_i=(N-\alpha)/2$ for $i=1,2,\ldots,N$,
but $F_0=N(\alpha+1)/2$.
\item
When $\beta=-2N-2-\alpha$ (with $-2-N<\alpha<-N$) one finds that all $F_i=-(\alpha+1)/2$ for $i=0,1,\ldots,N-1$,
but $F_N=N(N+\alpha+2)/2$.
\end{itemize}
For these cases, it could still be interesting to consider the Hamiltonian~\eqref{H} built from
the corresponding interaction matrix. The Hamiltonian, however, is not of the form~\eqref{H-gen}
as either the first or last oscillator in the chain would play a special role and give rise
to an extra term (either in $\q_1^2$ or else in $\q_n^2$); so the chain would no longer consist
of identical oscillators, but have one of them different from the others.

The Hamiltonians corresponding to $\beta=\alpha$ will now be considered in the next subsection.

\subsection{Hamiltonian with Hahn interaction}

Consider a linear chain of $n$ identical harmonic oscillators with a nearest neighbour 
interaction given by
\begin{equation}
\hH_Q=\sum_{r=1}^{n} \Big( \frac{\hat{p}_r^2}{2m}
+ \frac{m\omega^2}{2} \hat{q}_r^2 \Big) - \frac{cm}{2} \sum_{r=1}^{n-1} 
\sqrt{\frac{r(n-r)(r+2\alpha)(r+2\alpha+n)}{(2r+2\alpha-1)(2r+2\alpha+1)}}\;
\hat{q}_r\hat{q}_{r+1} ,
\label{HQ1}
\end{equation}
the interaction term to be referred to as the ``Hahn interaction'',
and where $\alpha$ is some parameter with $\alpha>-1$ or $\alpha<-n+1$
(guaranteeing that the expression under the square root is positive).
This Hamiltonian can be written in matrix form:
\begin{equation}
\hH_Q=\frac{1}{2m} \left(\begin{array}{ccc} \p_1^\dagger & \cdots & \p_n^\dagger \end{array}\right)
\left(\begin{array}{c} \p_1 \\ \vdots \\ \p_n \end{array}\right) + 
\frac{m}{2} \left(\begin{array}{ccc} \q_1^\dagger & \cdots & \q_n^\dagger \end{array}\right) 
\big((\omega^2-\frac{c(n-1)}{2}) I + c M_Q\big) 
\left(\begin{array}{c} \q_1 \\ \vdots \\ \q_n \end{array}\right),
\label{HQm1}
\end{equation}
where $M_Q$ is the matrix~\eqref{MQ} with $\beta=\alpha$ and $N=n-1$.
Since the diagonal matrix $D$ in the spectral decomposition of $M_Q$ is again 
$\hbox{diag}(0,1,2\ldots,n-1)$, it follows that:
\begin{prop}
The Hamiltonian $\hH_Q$ given by \eqref{HQ1} is analytically solvable.
The spectrum of $\hH_Q$ is exactly
the same as that of $\hH_K$, and given by~\eqref{specHK} and~\eqref{omegaK}.
\end{prop}
Also the condition for positive definiteness is the same as in the Krawtchouk case, 
namely $c<2\omega^2/(n-1)$.
Note that for $\alpha=1/2$ the form of the interaction
term is considerably simpler:
\begin{equation}
\hat{H}_Q^{(1)}=\sum_{r=1}^{n} \Big( \frac{\hat{p}_r^2}{2m}
+ \frac{m\omega^2}{2} \hat{q}_r^2 \Big) - \frac{cm}{2} \sum_{r=1}^{n-1} 
\frac{\sqrt{(n-r)(n+r+1)}}{2}\;
\hat{q}_r\hat{q}_{r+1}.
\label{HC1}
\end{equation}
Observe that for $\alpha\rightarrow +\infty$, the interaction~\eqref{HQ1} reduces to the
Krawtchouk interaction~\eqref{HK}.
Also note that under the reflection $r\rightarrow n-r$ in the interaction term, the Hamiltonian
reads
\begin{equation}
\hat{H}_Q^{(2)}=\sum_{r=1}^{n} \Big( \frac{\hat{p}_r^2}{2m}
+ \frac{m\omega^2}{2} \hat{q}_r^2 \Big) - \frac{cm}{2} \sum_{r=1}^{n-1} 
\frac{\sqrt{r(2n-r+1)}}{2}\;
\hat{q}_r\hat{q}_{r+1}.
\label{HC2}
\end{equation}
This can also be obtained by taking $\alpha=-n-1/2$ in~\eqref{HQ1}.

\section{$q$-Krawtchouk interaction} \label{sec5}

\subsection{The dual $q$-Krawtchouk polynomials}
For a fixed positive integer parameter $N$, and real parameters $q>0$ 
and $\bar c<0$~\footnote{In standard literature, this parameter is usually denoted by $c$, but we 
replace it by $\bar c$ in order not to confuse with the notation for the coupling constant $c$.}, 
the dual $q$-Krawtchouk
polynomial of degree~$i$ in the variable $\lambda(x)=q^{-x}+\bar c q^{x-N}$ is defined by~\cite{Koekoek,Ismail}
\begin{equation}
K_i(\lambda(x);q) \equiv K_i(\lambda(x);\bar c,N;q) = 
{\;}_3\phi_2 \left( \atop{q^{-i}, q^{-x}, \bar c q^{x-N}}{q^{-N},0} ; q;q \right),
\label{defKq}
\end{equation}
where ${\;}_3\phi_2$ is the $q$-generalized (or basic) hypergeometric series (which is terminating here due to the
numerator parameter $q^{-i}$). Recall that~\cite{Slater}
\[
{\;}_3\phi_2 \left( \atop{a,b,c}{d,e} ; q;z \right) =
\sum_{k=0}^\infty \frac{(a;q)_k(b;q)_k(c;q)_k}{(d;q)_k(e;q)_k(q;q)_k}z^k,
\]
where the $q$-Pochhammer symbol is~\cite{Slater}
\[
(a;q)_n = (1-a)(1-aq)\cdots (1-aq^{n-1}).
\]
The dual $q$-Krawtchouk polynomials satisfy the discrete orthogonality relation:
\begin{equation}
\sum_{x=0}^N w(x) K_i(\lambda(x);q) K_j(\lambda(x);q) = h_i \delta_{ij},
\label{orth-Kq}
\end{equation} 
where
\begin{align*}
w(x) &= \frac{(\bar cq^{-N};q)_x(q^{-N};q)_x (1-\bar cq^{2x-N})}{(q;q)_x(\bar cq;q)_x(1-\bar cq^{-N})}
 \bar c^{(-x)}q^{x(2N-x)} \quad (x=0,1,\ldots,N); \\
h_i &= (\bar{c}^{-1};q)_N \frac{(q;q)_i}{(q^{-N};q)_i}(\bar cq^{-N})^i .
\end{align*}
The recurrence relation for dual $q$-Krawtchouk polynomials is given by
\begin{align}
&-(1-q^{-x})(1-\bar cq^{x-N})K_i(\lambda(x);q) = 
\bar c q^{-N}(1-q^{i})K_{i-1}(\lambda(x);q) \nn\\
&\qquad -[(1-q^{i-N})+\bar cq^{-N}(1-q^{i})]K_i(\lambda(x);q)
+(1-q^{i-N})K_{i+1}(\lambda(x);q) 
\label{Kq-rec1}
\end{align}
The orthonormal dual $q$-Krawtchouk polynomials are defined by
\begin{equation}
\label{Kq-orth}
\Tilde K_i(\lambda(x);q) = \frac{\sqrt{w(x)} \, K_i(\lambda(x);q)}{\sqrt{h_i}},  \qquad i = 0,1,2, \ldots, N.
\end{equation}
Then we have the usual lemma:
\begin{lemm} \label{Kq-lemma}
Let $M_{Kq}$ be the tridiagonal $(N+1)\times(N+1)$-matrix
\begin{equation}
\label{MKq}
M_{Kq}= \left( \begin{array}{ccccc}
             F_0 & -E_1  &    0   &        &      \\
            -E_1 &  F_1  &  -E_2  & \ddots &      \\
              0  & -E_2  &   F_2  & \ddots &  0   \\
                 &\ddots & \ddots & \ddots & -E_N \\
                 &       &    0   &  -E_N  &  F_N
          \end{array} \right),
\end{equation}
where
\begin{equation} 
\label{Kq_EF}
E_i =\sqrt{\bar cq^{-N}(1-q^i)(1-q^{i-1-N})}
\quad
F_i=(1-q^{i-N})+\bar cq^{-N}(1-q^{i}),
\end{equation}
and let $U$ be the $(N+1)\times(N+1)$-matrix with matrix elements
\begin{equation}
U_{ij} = \Tilde K_i(\lambda(j);q) 
\end{equation}
where $i,j=0,1,\ldots,N$. 
Then
\begin{equation}
UU^T = U^TU=I \qquad\hbox{and} \qquad M_{Kq} = UDU^T
\end{equation}
where $D = \hbox{diag}((1-q^{-j})(1-\bar cq^{j-N}))$, $(j=0,1,2\ldots,N)$.
\end{lemm}

The proof is the same as those given before, and uses the orthonormality of $\Tilde K_i(\lambda(j);q)$ and
\begin{equation}
\label{Kq_rec2}
(1-q^{-x})(1-\bar cq^{x-N}) \Tilde K_i(\lambda(x);q) = - E_i \,\Tilde K_{i-1}(\lambda(x);q)
           + F_i \,\Tilde K_i(\lambda(x);q) 
           - E_{i+1} \,\Tilde K_{i+1}(\lambda(x);q).
\end{equation}

This case is interesting because for $\bar c=-1$ the 
diagonal of~\eqref{MKq} is constant, in other words then $F_i$ is 
independent of~$i$. 
This leads again to a Hamiltonian of the type~\eqref{H-gen}.

\subsection{Hamiltonian with dual $q$-Krawtchouk interaction}

Now we consider a linear chain of $n$ identical harmonic oscillators with a nearest neighbour 
interaction given by
\begin{equation}
\hH_{Kq}=\sum_{r=1}^{n} \Big( \frac{\hat{p}_r^2}{2m}
+ \frac{m\omega^2}{2} \hat{q}_r^2 \Big) - \frac{cm}{2} \sum_{r=1}^{n-1} 
2\sqrt{q^{r+1-2n}(1-q^r)(1-q^{n-r})}\;
\hat{q}_r\hat{q}_{r+1} ,
\label{HKq}
\end{equation}
the interaction term to be referred to as ``dual $q$-Krawtchouk interaction'',
where $q>0$.
This Hamiltonian can be written in matrix form:
\begin{equation}
\hH_{Kq}=\frac{1}{2m} \left(\begin{array}{ccc} \p_1^\dagger & \cdots & \p_n^\dagger \end{array}\right)
\left(\begin{array}{c} \p_1 \\ \vdots \\ \p_n \end{array}\right) + 
\frac{m}{2} \left(\begin{array}{ccc} \q_1^\dagger & \cdots & \q_n^\dagger \end{array}\right) 
\big((\omega^2-c(1-q^{1-n})) I + c M_{Kq}\big) 
\left(\begin{array}{c} \q_1 \\ \vdots \\ \q_n \end{array}\right),
\label{HKq1}
\end{equation}
where $M_{Kq}$ is the matrix~\eqref{MKq} with $\bar c=-1$ and $N=n-1$.
It follows that the spectrum of $\hH_{Kq}$ is given by~\eqref{specHK}, with
\begin{equation}
\omega_j = \sqrt{ \omega^2-c(1-q^{1-n}) + c(1-q^{-j})(1+q^{j-n+1})  } = 
\sqrt{ \omega^2+c(q^{j-n+1}-q^{-j})}, \qquad (j=1,2,\ldots,n).
\label{q-omega-j}
\end{equation}
For positive definiteness of the interaction matrix, all 
quantities under the square root must be positive.
It is easy to see that $\omega^2+c(q^{j-n+1}-q^{-j})$ ($j=1,2,\ldots,n$) is an increasing
sequence of $j$ when $q>1$ and a decreasing sequence of $j$ when $q<1$.
So for $q>1$ the condition means $c<q\omega^2/(1-q^{3-n})$,
while for $0<q<1$ we need $c<q^n\omega^2/(1-q^{n+1})$.
To conclude, we have
\begin{prop}
The Hamiltonian $\hH_{Kq}$ given by~\eqref{HKq} is analytically solvable.
The explicit spectrum of $\hH_{Kq}$ is given by~\eqref{specH},
where the constants $\omega_j$ are given by $\omega_j =\sqrt{\omega^2+c(q^{j-n+1}-q^{-j})}$.
\end{prop}

\section{Some properties of the spectra and conclusion} \label{sec6}

The spectrum of each of the Hamiltonians given here is of the form~\eqref{specH}, thus it is discrete but
infinite dimensional.
In order to appreciate the differences of the various examples given here, we shall plot the
energy levels of the singly excited states (the single phonons, or the simple vibrations) of the system.
These are the levels of the $n$ states $|1,0,\ldots,0\rangle$, $|0,1,0,\ldots,0\rangle$, 
$\ldots,$ $|0,\ldots,0,1\rangle$ (in the notation of~\eqref{specH}).
So, following~\eqref{specH}, these levels are given by
\begin{equation}
E_0+\hbar\omega_1, E_0+\hbar\omega_2,\ldots, E_0+\hbar\omega_n, 
\end{equation}
where
\begin{equation}
E_0= \frac{1}{2} \sum_{j=1}^n \hbar\omega_j.
\end{equation}
In order to illustrate the spacing of the energy levels of the singly excited states, it is sufficient
to plot the values of $(\omega_1,\omega_2,\ldots,\omega_n)$. We plot these values in 
Figure~1, for $n=12$, in four different cases:
\begin{itemize}
\item[(a)] The Hamiltonian~\eqref{H1-tilde} with constant nearest neighbour interaction,
where the values $\omega_j$ are given by~\eqref{omega-a}, plotted in Figure~1(a).
\item[(b)] The Hamiltonian with Krawtchouk interaction~\eqref{HK} or with Hahn interaction~\eqref{HQ1},
which have the same spectrum and where the values $\omega_j$ are given by~\eqref{omegaK}, plotted in Figure~1(b).
\item[(c)] The Hamiltonian with $q$-Krawtchouk interaction~\eqref{HKq} where $q>1$, for which the 
$\omega_j$ are given by~\eqref{q-omega-j}, plotted in Figure~1(c).
\item[(d)] The same Hamiltonian~\eqref{HKq} but with $q<1$, for which the 
$\omega_j$ are also given by~\eqref{q-omega-j}, plotted in Figure~1(d).
\end{itemize}
The values of $c$ and $q$ are appropriately chosen (see the figure caption for actual values) in order to
illustrate the typical energy level spacing properties for each case.

For a Hamiltonian with constant nearest neighbour interaction like~\eqref{H1} or~\eqref{H1-tilde},
the levels are wider apart in the middle of the spectrum, and closer to each other near the top and
the bottom of the spectrum, see Figure~1(a). The property is known, and was e.g.\ also observed
in~\cite{Iachello}.
For a Hamiltonian with a Krawtchouk interaction or a Hahn interaction like~\eqref{HK} or~\eqref{HQ1},
the energy level spacing decreases as the energy increases, a phenomenon also typical for
molecular spectra, see Figure~1(b).
Finally, for a Hamiltonian with a $q$-Krawtchouk interaction like~\eqref{HKq},
the energy level properties depend on whether $0<q<1$ or $q>1$. 
For $q>1$, one observes just the opposite of a constant nearest neighbour interaction:
the energy level spacing is small near the middle of the spectrum, and larger near the top
and bottom of the spectrum (but they are wider apart near the top than near the bottom), see Figure~1(c). 
For $0<q<1$, the levels behave similarly:
the energy level spacing is small near the middle of the spectrum, and larger near the top
and bottom of the spectrum, but now they are wider apart near the bottom than near the top, see Figure~1(d). 
It should be noted that for the cases (a), (b) and (c) the order of the levels from bottom
to top correspond to the states $|1,0,\ldots,0\rangle$, $|0,1,\ldots,0\rangle$, $\ldots,$ $|0,0,\ldots,1\rangle$ 
in this order, whereas for (d) it is just the opposite. This is related to the fact that
the sequence $\omega_j$ ($j=1,2,\ldots,n$) is an increasing sequence in the cases (a), (b) and (c), but
a decreasing sequence in the case (d) (see the remark following eq.~\eqref{q-omega-j}).

To conclude, we have presented a number of Hamiltonians for a quantum system consisting of a linear chain of
identical oscillators with a nearest neighbour interaction term depending on the position in the chain,
in general given by~\eqref{H-gen}.
Although such Hamiltonians are always numerically solvable, we have focused on the question
when such a Hamiltonian is analytically solvable.
The connection between the interaction matrix of the matrix expression of the Hamiltonian and 
Jacobi matrices of discrete orthogonal polynomials has led to new analytically solvable Hamiltonians,
with interesting spectral properties.

Our original interest in Hamiltonians with a nearest neighbour interaction of the form~\eqref{H1}
stems from the fact that this Hamiltonian can also be quantized in an alternative way,
namely as a Wigner quantum system~\cite{Wigner,Palev86,Palev1}, 
leading in particular to finite spectra and
non-commutative coordinates~\cite{PS1,PS2,KPSV1,KPSV2,LSV06,LSV08}. 
Our aim is to investigate when Hamiltonians with a general interaction matrix,
of the form~\eqref{H}, can still be solved as a Wigner quantum system, and
investigate the properties of such a solution. We hope to report on that
topic in the near future.

\newpage
\begin{figure}[htb]
\caption{Energy levels of the $n$ single phonon states $|0,\ldots,0,1,0,\ldots,0\rangle$, for $n=12$.
In each of the cases, $\hbar=\omega=1$. The four cases correspond to:
(a) a Hamiltonian~\eqref{H1-tilde} with constant nearest neighbour interaction [for $c=0.5$];
(b) a Hamiltonian with Krawtchouk interaction~\eqref{HK} or with Hahn interaction~\eqref{HQ1}
[for $c=0.18$];
(c) a Hamiltonian with $q$-Krawtchouk interaction~\eqref{HKq} where $q>1$ [here, $q=1.6$ and $c=1.0$];
(d) the same $q$-Krawtchouk interaction~\eqref{HKq} where $0<q<1$ [here, $q=0.7$ and $c=0.01$].
The levels are rescaled, so that the lowest and highest levels match in the four cases.} 
\[
\begin{tabular}{ccccccc}
(a) && (b) && (c) && (d) \\[5mm]
\includegraphics{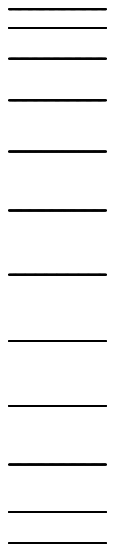} && \includegraphics{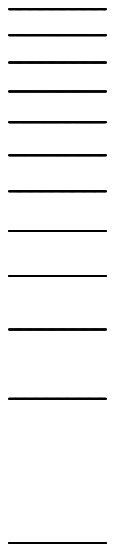} &&\includegraphics{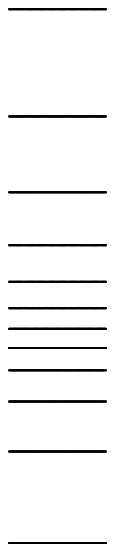} &&\includegraphics{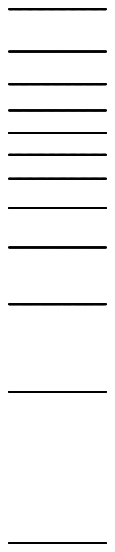}
\end{tabular}
\]
\end{figure}
\newpage

\section*{Acknowledgments}
G.\ Regniers was supported by project P6/02 of the Interuniversity Attraction Poles Programme (Belgian State -- 
Belgian Science Policy).


\begin{thebibliography}{99}
\bibitem{Schrodinger}
E.\ Schr\"odinger,
Ann.\ der Physik {\bf 349}(14), 916-934 (1914).
\bibitem{Christian}
W.G.\ Christian, A.G.\ Law, W.F.\ Martens, A.L.\ Mullikin and M.B.\ Sledd,
J.\ Math.\ Phys. {\bf 17}, 146-158 (1976).
\bibitem{Law}
A.G.\ Law and M.B.\ Sledd,
Lect.\ Notes Math.\ {\bf 1171}, 506-513 (1985).
\bibitem{Mokole}
E.L.\ Mokole, A.L.\ Mullikin and M.B.\ Sledd,
J.\ Math.\ Phys.\ {\bf 31}, 1902-1913 (1990).
\bibitem{Toda}
M.\ Toda,
Prod.\ Jpn.\ Acad., Ser.\ B {\bf 80}, 445-458 (2004).
\bibitem{Calogero}
F.\ Calogero,
J.\ Math.\ Phys. {\bf 12}, 419-436 (1971).
\bibitem{Sutherland}
B.\ Sutherland,
Phys.\ Rev.\ A {\bf 5}, 1372-1376 (1972).
\bibitem{Moser}
J.\ Moser,
Adv.\ Math. {\bf 16}, 197-220 (1975).
\bibitem{Olshanetsky}
M.A.\ Olshanetsky and A.M.\ Perelomov,
Lett.\ Math.\ Phys. {\bf 2}, 7-13 (1977).
\bibitem{Corrigan}
E.\ Corrigan and R.\ Sasaki,
J.\ Phys.\ A: Math.\ Gen. {\bf 35}, 7017-7061 (2002).
\bibitem{Karimipour}
V.\ Karimipour,
J.\ Math.\ Phys. {\bf 38}, 1577-1582 (1997).
\bibitem{Jain}
S.R.\ Jain and A.\ Khare,
Phys.\ Lett.\ A {\bf 262}, 35-39 (1999).
\bibitem{Basu}
B.\ Basu-Mallik and A.\ Kundu,
Phys.\ Lett.\ A {\bf 279}, 29-32 (2001).
\bibitem{Auberson}
G.\ Auberson, S.R.\ Jain and A.\ Khare,
J.\ Phys.\ A: Math.\ Gen. {\bf 34}, 695-724 (2001).
\bibitem{Cramer}
M.\ Cramer and J.\ Eisert,
New J.\ Phys. {\bf 8}, 71 (2006).
\bibitem{Amico}
L.\ Amico, R.\ Fazio, A.\ Osterloh and V.\ Vedral,
Rev.\ Mod.\ Phys. {\bf 80}, 517-576 (2008).
\bibitem{Plenio}
M.B.\ Plenio, J.\ Hartley and J.\ Eisert, 
New J.\ Phys. {\bf 6}, 36 (2004).
\bibitem{Cohen}
C.\ Cohen-Tannoudji, B.\ Diu  and F.\ Lalo\"e,
{\em Quantum Mechanics} (Wiley,  New York, 1977),
Vol.~1, complement JV.
\bibitem{LSV06}
S.\ Lievens, N.I.\ Stoilova and J.\ Van der Jeugt, 
J.\ Math.\ Phys. {\bf 47}, 113504 (2006).
\bibitem{LSV08}
S.\ Lievens, N.I.\ Stoilova and J.\ Van der Jeugt,
J.\ Math.\ Phys.\ {\bf 49}, 073502 (2008).
\bibitem{Iachello}
F.\ Iachello and A.\ Del Sol Mesa,
J.\ Math.\ Chem. {\bf 25}, 345-363 (1999).
\bibitem{Golub}
G.H.\ Golub and C.F.\ Van Loan, 
{\em Matrix Computations} (Johns Hopkins University Press, Baltimore, 1996).
\bibitem{Moshinsky}
M.\ Moshinsky, 
{\em The Harmonic Oscillator in Modern Physics: from Atoms to Quarks}
(Gordon and Breach, New-York, 1969).
\bibitem{Wybourne}
B.G.\ Wybourne, 
{\em Classical Groups for Physicists} 
(Wiley, New York, 1974).
\bibitem{Koekoek}
R.\ Koekoek and R.F.\ Swarttouw
{\em The Askey-scheme of hypergeometric orthogonal polynomials and its
  $q$-analogue}
(Technical Report 98--17, Delft University of Technology, 1998).
\bibitem{Ismail}
M.E.H.\ Ismail,
{\em Classical and quantum orthogonal polynomials in one variable}
(Cambridge University Press, 2005).
\bibitem{Suslov}
A.F.\ Nikiforov, S.K.\ Suslov and V.B.\ Uvarov, 
{\em Classical Orthogonal Polynomials of a Discrete Variable}
(Springer-Verlag, Berlin, 1991).
\bibitem{Slater}
L.J.\ Slater,
{\em Generalized Hypergeometric Functions}
(Cambridge University Press, 1966).
\bibitem{Wigner}
E. P.\ Wigner,  
Phys.\ Rev. {\bf 77}, 711-712 (1950).
\bibitem{Palev86}
A.H.\ Kamupingene, T.D.\ Palev and S.P.\ Tsavena, 
J.\ Math.\ Phys. {\bf 27}, 2067-2075 (1986).
\bibitem{Palev1}
T.D.\ Palev, 
J.\ Math.\ Phys. {\bf 23}, 1778-1784 (1982);   
Czech.\ J.\ Phys., Sect. B {\bf 29}, 91-98 (1979). 
\bibitem{PS1}
T.D.\ Palev and N.I.\ Stoilova, 
J.\ Math.\ Phys. {\bf 38}, 2506-2523 (1997).
\bibitem{PS2}
T.D.\ Palev and N.I.\ Stoilova, 
J.\ Phys.\ A: Math.\ Gen. {\bf 27}, 7387-7401 (1994).
\bibitem{KPSV1}
R.C.\ King, T.D.\ Palev, N.I.\ Stoilova and J.\ Van der Jeugt,  
J.\ Phys.\ A: Math. Gen. {\bf 36} 
4337-4362 (2003).
\bibitem{KPSV2}
R.C.\ King, T.D.\ Palev, N.I.\ Stoilova  and J. Van der Jeugt, 
J.\ Phys.\ A: Math.\ Gen. {\bf 36} 11999-12019
(2003).
\end{thebibliography}
\end{document}